\documentclass[amsmath,amssymb,twocolumn,showpacs,prl,superscriptaddress]{revtex4}
\usepackage{graphicx}

\begin{document}

\title{Evidence for entangled states of two coupled flux qubits}

\author{A.~Izmalkov}
\affiliation{%
Institute for Physical High Technology, P.O. Box 100239, D-07702
Jena, Germany}
\affiliation{%
Moscow Engineering Physics Institute (State University),
Kashirskoe shosse 31, 115409 Moscow, Russia}

\author{M.~Grajcar}
\affiliation{%
Institute for Physical High Technology, P.O. Box 100239, D-07702
Jena, Germany} \affiliation {Department of Solid State Physics,
Comenius University, SK-84248 Bratislava, Slovakia}

\author{E.~Il'ichev}
\email{ilichev@ipht-jena.de}
\affiliation{%
Institute for Physical High Technology, P.O. Box 100239, D-07702
Jena, Germany}

\author{Th.~Wagner}
\affiliation{%
Institute for Physical High Technology, P.O. Box 100239, D-07702
Jena, Germany}

\author{H.-G.~Meyer}
\affiliation{%
Institute for Physical High Technology, P.O. Box 100239, D-07702
Jena, Germany}

\author{A.Yu.~Smirnov}
\affiliation{%
D-Wave Systems Inc., 320-1985 W. Broadway, Vancouver, B.C., V6J
4Y3 Canada}

\author{M.H.S.~Amin}
\affiliation{%
D-Wave Systems Inc., 320-1985 W. Broadway, Vancouver, B.C., V6J
4Y3 Canada}

\author{Alec Maassen van den Brink}
\affiliation{%
D-Wave Systems Inc., 320-1985 W. Broadway, Vancouver, B.C., V6J
4Y3 Canada}

\author{A.M.~Zagoskin}
\email{zagoskin@dwavesys.com}
\affiliation{%
D-Wave Systems Inc., 320-1985 W. Broadway, Vancouver, B.C., V6J
4Y3 Canada}
\affiliation{%
Physics and Astronomy Dept., The University of British Columbia,
6224 Agricultural Rd., Vancouver, B.C., V6T 1Z1 Canada}

\date{\today}

\pacs{85.25.Cp
, 03.67.Lx
, 03.67.-a
, 85.25.Dq}

\begin{abstract}
We have studied the low-frequency magnetic susceptibility of two inductively coupled flux qubits using the impedance measurement technique (IMT), through their influence on the resonant properties of a weakly coupled high-quality tank circuit. In a single qubit, an IMT dip in the tank's current--voltage phase angle at the level anticrossing yields the amplitude of coherent flux tunneling. For two qubits, the difference (IMT deficit) between the sum of single-qubit dips and the dip amplitude when both qubits are at degeneracy shows that the system is in a mixture of entangled states (a necessary condition for entanglement). The dependence on temperature and relative bias between the qubits allows one to determine all the parameters of the effective Hamiltonian and equilibrium density matrix, and confirms the formation of entangled eigenstates.
\end{abstract}
\maketitle

While its exact role in the power of quantum computers can still be debated, there is no doubt that entanglement is a key feature of quantum registers, absent in their classical counterparts~\cite{N&C}. From the point of view of scalability, implementing quantum registers as integrated circuits is attractive. However, demonstrating entanglement in macroscopic solid-state systems remains a daunting task, even after the observation of quantum coherence in many types of superconducting qubits \cite{Nakamura99,Lukens,VdW,Martinis,Vion} and also entanglement in charge~\cite{Pashkin} and current-biased Josephson-junction~\cite{Berkley} qubits, using pulse and spectroscopic techniques respectively. In this paper we investigate two coupled three-junction flux qubits~\cite{Mooij,Majer} using a different method, providing a simple criterion for eigenstate entanglement, by studying the latter's influence on the two-qubit magnetic susceptibility.

The impedance measurement technique (IMT)~\cite{Greenberg:2002} relies on monitoring the
current and voltage in a high-quality, low-frequency resonant tank circuit inductively coupled to the qubit. In various modifications, this method was applied to map the Josephson potential profile~\cite{APL_IMT}, and to observe Landau--Zener transitions~\cite{Landau-Zener} and Rabi oscillations~\cite{Rabi} in a three-junction flux qubit. Due to the tank's weak coupling to the qubit and high quality factor, decoherence times can be as high as 2.5~$\mu$s~\cite{Rabi}. Note that the IMT approach requires a small decoherence rate compared to only the tunneling amplitudes (0.5$\sim$1~GHz), not the tank frequency (10$\sim$20~MHz). This paper generalizes to two qubits an IMT study of coherent tunneling~\cite{IMT:2003}. The tank is fed a small-amplitude ac signal; its effective impedance, and therefore the current--voltage phase angle, are sensitive to the qubit susceptibility~$\chi$. For, say, one qubit at temperature~$T=0$, $\chi\propto\partial^2_\Phi E_-$, the curvature of the ground-state energy~$E_-$ vs external flux~$\Phi$ [a different expression for $\chi$ is used in Eq.~(\ref{eq_suscept}) below]. Thus, the level anticrossing due to flux tunneling is revealed by a large peak in~$|\chi|$.

\begin{figure}
\includegraphics[width=3in]{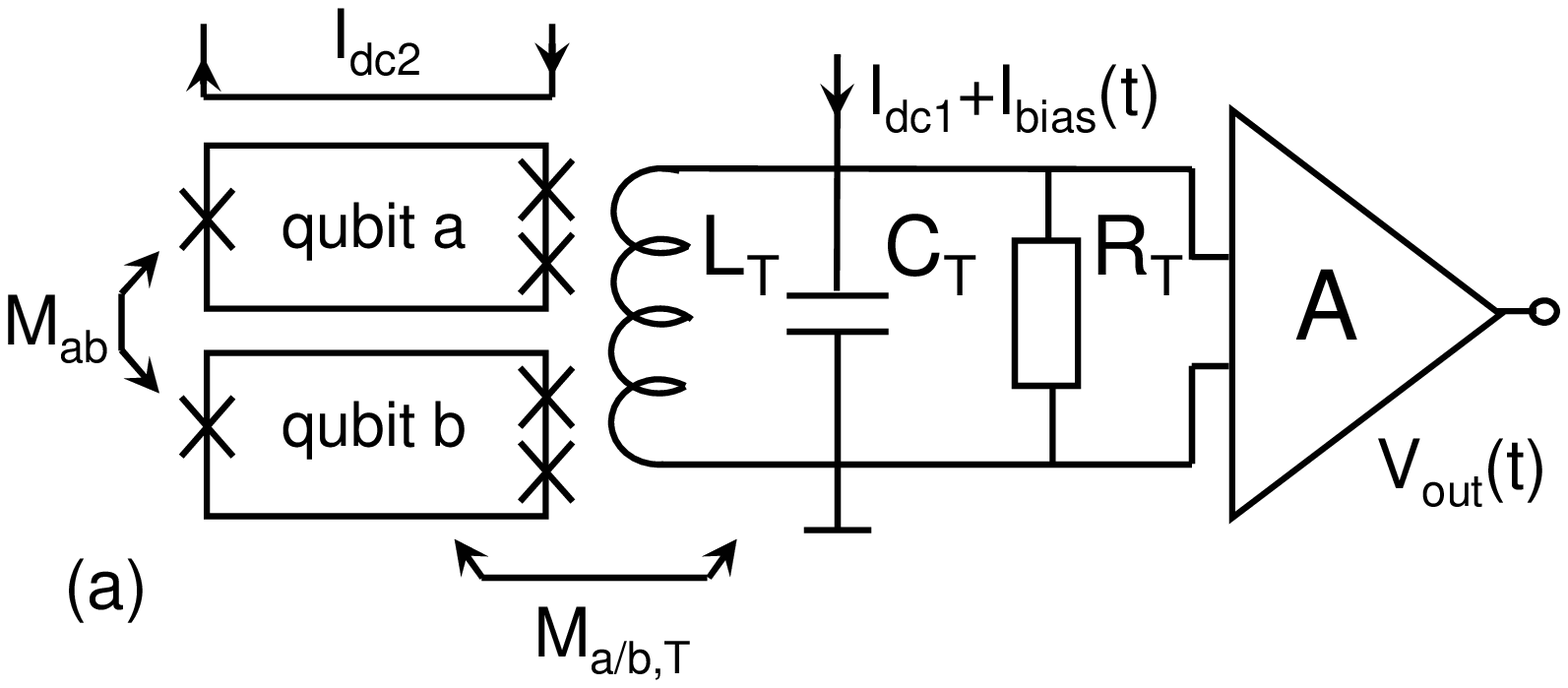}
\includegraphics[width=3in]{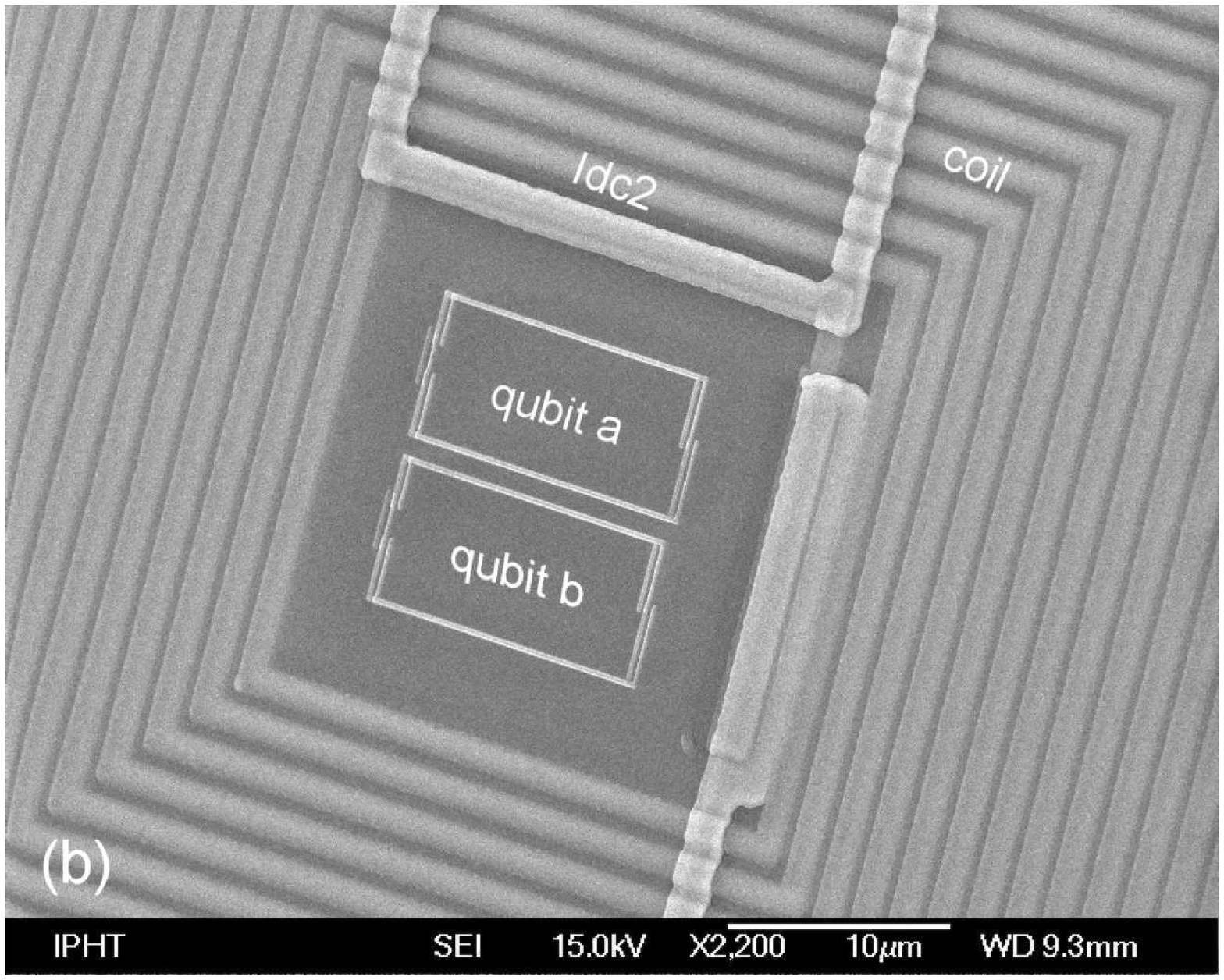}
\caption{Two-qubit system coupled to a resonant tank circuit. (a)
Schematic. (b) Micrograph.} \label{fig1}
\end{figure}

Our system of two flux qubits inductively coupled to each other and to the tank is shown in Fig.~\ref{fig1}. The qubits were fabricated out of aluminum by conventional shadow evaporation, nominally 1~$\mu$m apart, at the centre of a niobium pickup coil. The area of each qubit was $80$~$\mu$m$^2$, with self-inductance $L_{a/b}\approx39$~pH; the two biggest junctions in each qubit had critical current $\approx 400$~nA and Coulomb energy $e^2\!/2Ch\approx3.2$~GHz (for the third junction, these values are 10$\sim$20\% smaller and larger, respectively). The mutual inductance between the qubits $M_{ab}=2.7$~pH was estimated numerically from the electron micrograph. The Nb coil, which together with an external capacitor of $C_\mathrm{T}\approx470$~pF forms a parallel tank circuit with inductance $L_\mathrm{T}\approx130$~nH, resonant frequency
$\omega_\mathrm{T}/2\pi = 20.139$~MHz and quality factor $Q_\mathrm{T} \equiv\omega_\mathrm{T}R_\mathrm{T}C_\mathrm{T}=1680$ (at 10~mK; $R_\mathrm{T}$ is an effective tank resistance)~\cite{rectify}, was fabricated using e-beam lithography. The external magnetic flux through the qubits was created by the dc component of the current in the coil $I_\mathrm{dc1}$, and by the bias current $I_\mathrm{dc2}$ through a wire close to one of the qubits. This allowed independent control of the bias in each qubit.

The system  of Fig.~\ref{fig1} is described by the Hamiltonian $H = H_0 + H_\mathrm{T} + H_\mathrm{int} + H_\mathrm{diss}$, where the two-qubit Hamiltonian in the two-state approximation is expressed through Pauli matrices as~\cite{Alec03}
\begin{equation}\label{eq_Hamiltonian_0}
  H_0 =-\Delta_a\sigma_{x}^{(a)} - \Delta_b \sigma_{x}^{(b)} + \epsilon_a
  \sigma_{z}^{(a)} + \epsilon_b \sigma_{z}^{(b)} + J
  \sigma_{z}^{(a)} \sigma_{z}^{(b)},
\end{equation}
$H_\mathrm{T}$ is the tank Hamiltonian (a harmonic oscillator), the
qubit--tank interaction is
\begin{equation}\label{eq_Hamiltonian_int}
  H_\mathrm{int}=-(\lambda_a \sigma_{z}^{(a)}
  + \lambda_b \sigma_{z}^{(b)})I_\mathrm{T},
\end{equation}
and $H_\mathrm{diss}$ describes the standard weak coupling of the qubits to a dissipative bath~\cite{Weiss}. Here $I_\mathrm{T}$ is the current through~$L_\mathrm{T}$. The coefficients are $\lambda_{a/b} =  M_{a/b,\mathrm{T}} I_{a/b}$, where $M_{a/b,\mathrm{T}}$ is the qubit--tank mutual inductance and $I_{a/b}$ is the magnitude of the persistent
current in the corresponding qubit. The qubit biases are given by
$\epsilon_a = I_{a}\Phi_0 (f_\mathrm{x} - \frac{1}{2} + f_{\rm shift})$,
$\epsilon_b = I_{b}\Phi_0 (f_\mathrm{x} - \frac{1}{2} + \eta f_{\rm shift} )$, where $f_\mathrm{x}=\Phi/\Phi_0$ accounts for the external flux $\Phi\propto I_\mathrm{dc1}$ created by the Nb coil in both qubits, while the parameters $f_{\rm shift}\propto I_\mathrm{dc2}$ and $\eta = M_{b\mathrm{w}}/M_{a\mathrm{w}} < 1$ describe the bias difference between the qubits created by the additional wire. Here $M_{a\mathrm{w}}$ ($M_{b\mathrm{w}}$) are the mutual inductances between the $a$ ($b$) qubit and the additional dc wire (for our sample, $M_{a\mathrm{w}}$ and $M_{b\mathrm{w}}$ were calculated numerically, yielding $\eta=0.32$). The qubit--qubit coupling $J = M_{ab} I_{a}I_{b}$ is positive because the two qubits are in the same plane side by side, leading to antiferromagnetic coupling (according to the north-to-south attraction law).

The effective inductance of the tank, and therefore its resonant frequency, depend on the state of the qubits. In the IMT method~\cite{Greenberg:2002}, this dependence is observed by continuously monitoring the phase angle $\Theta$ between the average tank voltage and bias current $I_\mathrm{bias}(t)$ fed into the tank (Fig.~\ref{fig1}). For small enough $I_\mathrm{bias}$, $\tan\Theta$ is determined by the real part of the two-qubit contribution to the tank susceptibility $\chi'$ at~$\omega_\mathrm{T}$~\cite{SmirnovR2003}, viz.,
\begin{equation}\label{theta-chi}
  \tan \Theta = - \frac{Q_\mathrm{T}}{L_\mathrm{T}}\chi'(\omega_\mathrm{T}),
\end{equation}
where $\chi(\omega )$ has a Kubo-type linear-response expression through retarded Green's functions of the qubit operators~$\sigma_z^{(a/b)}$. For weak damping $H_\mathrm{diss}$, the latter can be calculated with the equilibrium density matrix corresponding to $H_0$ in Eq.~(\ref{eq_Hamiltonian_0}). It can be generally assumed that the eigenvalues $E_{\mu}$ of $H_0$, $\mu=\nobreak1,2,3,4$, are non-degenerate and the eigenstates normalized, $\langle \nu |\mu \rangle = \nobreak\delta_{\mu\nu}$. Taking into account the qubits' interaction with a dissipative environment \cite{SmirnovR2003,Blum}, a standard calculation gives~\cite{SmirnovUnpub}
\begin{equation}\label{eq_suscept}
 \chi(\omega ) = - \sum_{\mu\neq \nu}
 \frac{\rho_{\mu} -\rho_{\nu}}
 {\hbar\omega + E_{\mu} - E_{\nu} + i\hbar\Gamma_{\mu\nu}}R_{\mu \nu},
\end{equation}
where $\rho_{\mu} = e^{-E_{\mu}/T}/[\sum_{\nu}e^{-E_{\nu}/T}]$ is the population of energy level~$\mu$, $\Gamma_{\mu\nu}$ in the energy denominators are decoherence rates, and the real matrix elements are
\begin{eqnarray}\label{eq_tan_theta_R}
R_{\mu \nu} &=& \lambda_a^2 \langle \mu | \sigma_{z}^{(a)}|\nu
\rangle \langle \nu |\sigma_{z}^{(a)}|\mu \rangle
+ \lambda_b^2 \langle \mu |\sigma_{z}^{(b)}|\nu \rangle \langle \nu |\sigma_{z}^{(b)}|\mu \rangle\nonumber  \\
&&{}+ \lambda_a \lambda_b \langle \mu |\sigma_{z}^{(a)}|\nu
\rangle \langle \nu |\sigma_{z}^{(b)}|\mu \rangle\nonumber\\ &&{}+
\lambda_a \lambda_b \langle \mu |\sigma_{z}^{(b)}|\nu \rangle
\langle \nu |\sigma_{z}^{(a)}|\mu \rangle.
\end{eqnarray}
Substitution into Eq.~(\ref{theta-chi}) yields
\begin{equation}\label{eq_tan_theta}
  \tan \Theta = - 2\frac{Q_\mathrm{T}}{L_\mathrm{T}}  \sum_{\mu < \nu}\frac{
  \rho_{\mu} - \rho_{\nu}}{ E_{\nu} - E_{\mu} } R_{\mu \nu}.
\end{equation}
At low frequencies $\sim \omega_\mathrm{T} \ll |E_{\mu}{-}E_{\nu}|/\hbar$ and for weak damping $\Gamma_{\mu \nu}\ll |E_{\mu}{-}E_{\nu}|/\hbar$, the $\Gamma_{\mu\nu}$ do not affect $\tan\Theta$, but are responsible for establishing the equilibrium distribution.

The first two terms in Eq.~(\ref{eq_tan_theta_R}) are clearly positive, and non-zero even if the two-qubit states are factorized. The first (second) term corresponds to the contribution of qubit $a$ ($b$), and peaks near that qubit's degeneracy point. These contributions are practically independent of whether the qubits' degeneracy points coincide or not.

The last two terms in Eq.~(\ref{eq_tan_theta_R}) describe coherent flipping of both qubits, which is \emph{only} possible for non-factorizable (entangled) eigenstates $|\mu\rangle$,
$|\nu\rangle$. For $J>0$, these terms are found to be negative. Therefore the difference (\emph{IMT deficit}) between the coinciding two-qubit IMT dip and the sum of two single-qubit dips provides a measure of how coherent the two-qubit dynamics is, that is, whether entangled eigenstates of $H_0$ are formed~\cite{SmirnovUnpub}. Eigenstate entanglement is a necessary condition for the equilibrium state to be entangled; of course, the eigenstate populations play a role as well. However, it is already a sufficient condition for performing quantum gate operations (if decoherence times are long enough), since one can also initialize the system in a nonequilibrium state, say, by letting it relax at a point in parameter space for which there is a large gap above the ground state, and subsequently returning to the working point \cite{Nakamura99,Lukens}.

\begin{figure}
\includegraphics[width=3in]{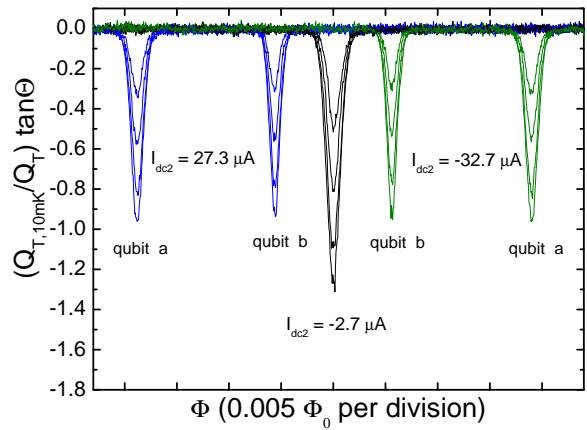}
\caption{Normalized tangent of the current--voltage angle in the
tank vs external bias~$\Phi$. From
the lower to upper curves, the temperature of the mixing chamber is
10, 50, 90, 160~mK. A relative flux bias $f_{\rm shift}$ between the
qubits is created by changing the current $I_\mathrm{dc2}$ in the
additional wire. The shifted curves correspond to $I_\mathrm{dc2} =
27.3$ ($-32.7$)~$\mu$A. Two IMT dips are then observed, showing
tunneling in each qubit at the corresponding degeneracy
point $f_\mathrm{x} \neq \frac{1}{2}$. The central curve has $I_\mathrm{dc2} = -2.7$~$\mu$A.
Both qubits are degenerate simultaneously at $f_\mathrm{x}=\frac{1}{2}$. One IMT
dip is observed, with an amplitude about 33\% (at small~$T$)
less than the sum of two separate dips.}
\label{fig2a}
\end{figure}

The measurements on the system of Fig.~\ref{fig1} allowed to determine all parameters of~$H_0$. The data are presented in Fig.~\ref{fig2a} ($\tan \Theta$ vs flux bias) and Fig.~\ref{fig3} ($T$-dependence of dip amplitudes).

\begin{figure}
\includegraphics[width=3in]{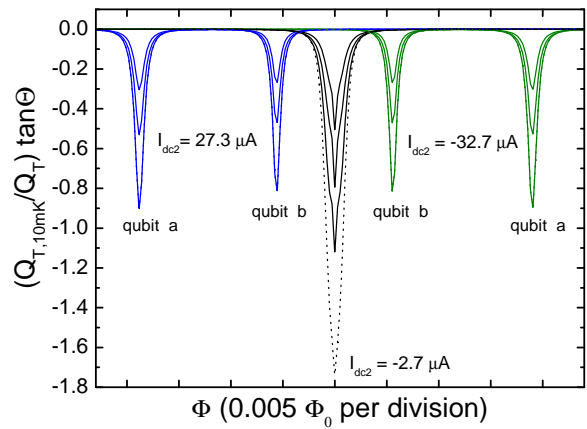}
\caption{Theoretical fit of the normalized tangent of the
current--voltage angle in the tank vs external
bias~$\Phi$. Solid curves, from lower to upper: $T=50$, 90, 160~mK. The interaction
energy between the qubits is $420$~MHz. The dashed central peak corresponds
to $T=50$~mK, but with the entangled contributions
in (\ref{eq_tan_theta_R}) omitted. Relative
flux biases are the same as for the experimental curves in
Fig.~\ref{fig2a}.} \label{fig2b}
\end{figure}

\begin{figure}
\includegraphics[width=3in]{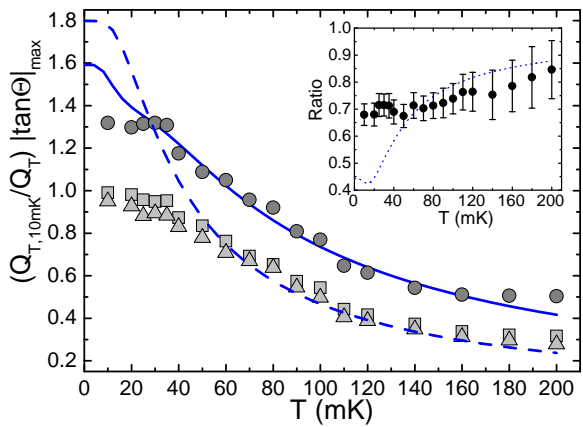}
\caption{Temperature dependence of the IMT dip amplitudes.
Squares: qubit~$a$; triangles: qubit~$b$; circles: coincident dip.
The solid and dashed curves are theoretical
fits (the dashed one is for qubit~$a$; the $b$-curve is essentially identical for $T>50$~mK). Inset: relative size of the coincident dip
$\tan\Theta_\mathrm{c}/\left(\tan\Theta_a{+}\tan\Theta_b\right)$; the
dotted line is the theoretical curve.} \label{fig3}
\end{figure}

First, consider $I_\mathrm{dc2} = 27.3$ ($-32.7$)~$\mu$A, i.e., unequally biased qubits. In this case, whenever one qubit is near-degenerate, the other is strongly biased, putting it into a unique classical ground state (at the low end of our $T$-range) in which it behaves trivially. Hence, one essentially observes single-qubit properties. We match the corresponding IMT dips of Fig.~\ref{fig2a} with the predictions of Eqs.~(\ref{eq_tan_theta}), (\ref{eq_tan_theta_R}) for both the width and amplitude; see Fig.~\ref{fig2b}. This yields $\Delta_a/h = 550$~MHz, $\Delta_b/h = 450$~MHz, and the persistent currents $I_a \approx I_b  \equiv I_\mathrm{p} = 320$~nA, for a flux-to-energy bias conversion factor $I_\mathrm{p}\Phi_0/h = 990$~GHz~\cite{Delta}. The width of the dips is practically $T$-independent, as expected~\cite{IMT:2003}. The $T$-dependence of the amplitudes (Fig.~\ref{fig3}, squares and triangles) agrees with these values for $\Delta_{a/b}$. (The saturation below 35--40~mK~\cite{IMT:2003} is likely due to a discrepancy between the mixing-chamber and sample temperatures.)

We can determine the coupling constant in Eq.~(\ref{eq_Hamiltonian_0}) from the \emph{single}-qubit measurements, as $J/h = M_{ab}I_\mathrm{p}^2/h = 410$~MHz. This agrees well with the value $J/h = 420$~MHz obtained by fitting the $T$-dependence of the coincident IMT dip amplitude above the saturation temperature (Fig.~\ref{fig3}, circles) with the theoretical curve.

Comparison of the single-qubit and coincident IMT dips shows that the contribution to $\tan \Theta$ due to eigenstate entanglement is significant. Indeed, the amplitude of the central dip in Fig.~\ref{fig2a} at $T=50$~mK is 1.12, compared to the sum 1.69 of the single-qubit dips. This means that the entangled terms [last two in Eq.~(\ref{eq_tan_theta_R})] are
responsible for a contribution $\approx-0.57$. As the dashed line in Fig.~\ref{fig2b} shows, dropping these terms would yield the coincident IMT dip in marked deviation from both the full theory and experiment.

The IMT deficit also confirms the sign of the qubit--qubit interaction: in the case of ferromagnetic coupling, the coincident dip should be \emph{larger} than the sum of the single-qubit ones.

Plotting the ratio of the coincident dip to the sum of single-qubit ones (Fig.~\ref{fig3}, inset), we see that it grows with~$T$, approaching~$\approx1$. This is to be expected: thermal excitations tend to destroy coherent correlations between the qubits (which will then behave as independent quantum systems with zero IMT deficit)---eventually, of course, destroying the single-qubit IMT dips as well.

At 50~mK, $T$ is comparable to the characteristic qubit energies (at the two-qubit degeneracy point, the top excited state is $\sim100$~mK above the ground state). The measurement time equals at least the tank saturation time $Q_\mathrm{T}/\omega_\mathrm{T}$ and is longer in practice, 1$\sim$10~ms. Since this exceeds any conceivable qubit relaxation time $\sim\Gamma_{\mu\nu}^{-1}$, the system has time to equilibrate. Indeed, the good quantitative agreement between experiment (Figs.~\ref{fig2a}, \ref{fig3}) and theory [Eq.~(\ref{eq_tan_theta})] in a wide range of $T$ confirms that our system is described by the equilibrium density matrix with the Hamiltonian~(\ref{eq_Hamiltonian_0}). In other words, it is in a mixture of entangled two-qubit eigenstates.

A measure of the entanglement of an arbitrary state is the concurrence $\mathcal{C}$~\cite{Wootters}, changing monotonically from 0 for non-entangled to 1 for maximally entangled states. For a normalized pure state $\alpha|00\rangle+\beta|01\rangle+\gamma|10\rangle+\delta|11\rangle$, one has $\mathcal{C}=2|\alpha\delta{-}\beta\gamma|$; in general, $\mathcal{C}$ depends both on the existence of entangled eigenstates and on their population, and can be erased by temperature faster than quantum coherence in individual qubits~\cite{Yu_Eberly}. Using the experimentally established parameters of the Hamiltonian (\ref{eq_Hamiltonian_0}), the concurrences of its eigenstates at the two-qubit degeneracy point are immediately calculated: $\mathcal{C}_1=\mathcal{C}_4 = 0.39$, $\mathcal{C}_2=\mathcal{C}_3 = 0.97$. For the equilibrium density matrix of our system at 10~mK, these give $\mathcal{C}_\mathrm{eq} = 0.33$. This value will decrease with rising~$T$, vanishing around 30~mK. However, from Eq.~(\ref{eq_tan_theta}) it can be seen that $\Theta$ depends on $T$ much more weakly (through the Boltzmann factors $\rho_\mu$), so that the IMT deficit can be reliably extracted even at temperatures where $\mathcal{C}_\mathrm{eq}$ is small or even zero.

We stress that for quantum computing, it is the existence of entangled eigenstates that
matters, not the measure of equilibrium entanglement [cf.\ the discussion a bit below Eq.~(\ref{eq_tan_theta})], since the system anyway must operate on a faster scale than the equilibration time. The crucial requirement is that the decoherence time $\tau_\mathrm{2q}$ of the entangled eigenstates exceeds the operation time. The presence of an IMT deficit shows that in our system $\tau_\mathrm{2q}$ is larger than $\min(\hbar/\Delta_{a/b},\hbar/J) \sim 1$~ns; its actual value can be determined using, e.g., the approach of~\cite{Rabi}.

In conclusion, we have investigated two inductively coupled aluminum flux qubits with independently controlled bias fluxes. The impedance measurement technique directly measures the low-frequency part of the system's magnetic susceptibility. The quantitative agreement between theory and experiment confirms that the system is in an equilibrium mixture of entangled states. Besides trying to scale the approach to larger qubit systems, it will next be important to study the system's coherence properties in detail, and to establish time-domain control enabling actual quantum gate operations.

\begin{acknowledgments}
We thank N. Oukhanski, H. M\"{u}hlig, and D. Born for technical assistance, U. H\"{u}bner and T. May for sample fabrication, and I.~Affleck, D. Bonn, M. Franz, W. Hardy, H.E. Hoenig, P.C.E. Stamp, and M.F.H. Steininger for fruitful discussions. MG acknowledges partial support by the Slovak Grant Agency VEGA (Grant No.\ 1/9177/02).
\end{acknowledgments}


\begin{thebibliography}{99}
\bibitem{rectify} Since the last factor in Eq.~(3) of Ref.~\cite{Rabi} has poles at $\omega\approx\pm\omega_\mathrm{T}-i\gamma_\mathrm{T}/2$, in that paper $\gamma_\mathrm{T}$ is the energy, not the amplitude, decay rate. Therefore, on p.~2, col.~2, l.~4--5, the definition should have read $Q_\mathrm{T}=\omega_\mathrm{T}/\gamma_\mathrm{T}$. However, the value $Q_\mathrm{T}=1850$ itself is correct.
\bibitem{Delta} In Fig.~\ref{fig2a}, the single-qubit dips having the larger dependence on $I_\mathrm{dc2}$ can be ascribed to qubit~$a$ because it is closest to the bias wire (Fig.~\ref{fig1}).
\end{thebibliography}
\end{document}